\documentclass[prd,nofootinbib,showpacs]{revtex4}
\usepackage{natbib}
\usepackage{amssymb,amsbsy,amsmath,amsfonts}
\usepackage{graphicx}
\usepackage{float}
\usepackage{rotating}

\def\slashp{p \!\!\! \slash}

\begin{document}
\title {Electromagnetic structure of the lowest-lying decuplet resonances in covariant chiral perturbation theory}

\author{L.S. Geng$^1$}
\author{J. Martin Camalich$^1$}
\author{M.J. Vicente Vacas$^1$}

\affiliation{$^1$Departamento de F\'{\i}sica Te\'orica and IFIC, Centro
Mixto, Institutos de Investigaci\'on de Paterna - Universidad de
Valencia-CSIC, Spain}

\begin{abstract}
We present a calculation of the leading SU(3)-breaking $\mathcal{O}(p^3)$-corrections to the electromagnetic moments and charge radius (CR) of the lowest-lying decuplet resonances in covariant chiral perturbation theory. In 
particular, the magnetic dipole moment (MDM) of the members of the decuplet is predicted fixing the only low-energy constant (LEC) present up to this order with the well measured MDM of the $\Omega^-$. We predict $\mu_\Delta^{++}=6.04(13)$ and $\mu_\Delta^+=2.84(2)$ which agree well with the current experimental information. For the electric quadrupole moment (EQM) and the CR we use state-of-the-art lattice QCD results to determine the corresponding LECs, whereas for the magnetic octupole moment (MOM) there is no unknown LEC up to the order considered here and we obtain a pure prediction. We compare our results with those reported in large $N_c$, lattice QCD, heavy-baryon chiral perturbation theory and other models.  
\end{abstract}
\date{\today}
\maketitle

\section{Introduction}

The $\Delta(1232)$ resonance is the lowest-lying excited state of the nucleon and plays a very important role in the low-energy baryon phenomenology. Unfortunately, its lifetime $\sim10^{-23}s$, marked by the strong decay into pion-nucleon, is too short and therefore its properties are only indirectly accessible in experiment. For instance, the electromagnetic form factors have been probed and the MDM of the $\Delta^{++}$ and of the $\Delta^{+}$ measured. In the former case, the radiative pion-nucleon scattering ($\pi^+p\longrightarrow\pi^+p\gamma$) is analyzed although the results of the different experiments~\cite{Nefkens:1977eb,Bosshard:1991zp,LopezCastro:2000cv} are not completely consistent. This is the reason behind the large uncertainties of the estimation quoted in the current Particle Data Group (PDG) review, $\mu_{\Delta^{++}}=3.7\sim7.5\,\mu_N$~\cite{Yao:2006px}. On the other hand, the magnetic dipole moment of $\Delta^{+}$ has been recently extracted from the radiative photo-production of neutral pions ($\gamma p\longrightarrow \pi^0 p\gamma'$), $\mu_{\Delta^{+}}=2.7^{+1.0}_{-1.3}\,({\rm stat})\,\pm\,1.5\,{\rm (syst)}\,\pm\,3\,{\rm (theor)}\,\mu_N$ ~\cite{Kotulla:2002cg}. A new experiment with the Crystal Ball detector at MAMI is expected to give soon new results with improved statistics~\cite{Kotulla:2008zz} and using theoretical extraction methods based either on a dynamical model~\cite{Chiang:2004pw} or on chiral effective field theory~\cite{Pascalutsa:2004je,Pascalutsa:2007wb}. Concerning the SU(3)-multiplet partners of the $\Delta(1232)$ resonances, namely the other members of the spin-3/2 lowest-lying decuplet, only the magnetic dipole moment of the $\Omega^-$ has been measured, $\mu_{\Omega^-}=-2.02\pm0.05\,\mu_N$~\cite{Yao:2006px}.

The electromagnetic properties of the decuplet resonances have been studied theoretically during the last two decades, and information not only on MDMs but also on other properties like the EQM, the MOM, or on the CR and the $q^2$ dependence of the form factors, have arisen from many different frameworks. Indeed, the electromagnetic structure of the decuplet baryons has been studied within the non relativistic quark model (NRQM) ~\cite{Hikasa:1992je,Krivoruchenko:1991pm}, the relativistic quark model (RQM)~\cite{Schlumpf:1993rm}, the chiral quark model ($\chi$QM)~\cite{Buchmann:1996bd,Wagner:2000ii}, the chiral quark soliton model ($\chi$QSM)~\cite{Kim:1997ip,Ledwig:2008es}, the spectator quark model (SpQM)~\cite{Ramalho:2008dc,Ramalho:2009vc}, the general parametrization method (GP)~\cite{Buchmann:2002xq,Buchmann:2008zza}, QCD sum rules (QCD-SR)~\cite{Lee:1997jk,Aliev:2000rc,Azizi:2008tx,Aliev:2009pd}, large $N_c$~\cite{Luty:1994ub,Jenkins:1994md,Buchmann:2002mm}, chiral perturbation theory ($\chi$PT)~\cite{Pascalutsa:2004je,Pascalutsa:2007wb,Butler:1993ej,Banerjee:1995wz,Arndt:2003we,Hacker:2006gu,Tiburzi:2009yd} and in lattice QCD (lQCD) ~\cite{Nozawa:1990gt,Leinweber:1992hy,Cloet:2003jm,Lee:2005ds,Alexandrou:2008bn,Aubin:2008qp,Alexandrou:2009hs,Boinepalli:2009sq}. Lately, the lQCD calculations have experienced a remarkable progress that allows a quantitative description of these properties from first principles. 

The $\chi$PT provides a model independent and systematic framework to study  the non-perturbative regime of the strong interactions~\cite{Gasser:1983yg,Gasser:1984gg,Gasser:1987rb,Scherer:2002tk}. The application of SU(3)-flavor $\chi$PT to the analysis of the electromagnetic properties of the decuplet, either in its full~\cite{Banerjee:1995wz,Butler:1993ej} or quenched versions~\cite{Arndt:2003we,Tiburzi:2009yd}, has been restrained to the heavy-baryon chiral perturbation theory (HB$\chi$PT) approach~\cite{Jenkins:1990jv}. Recently, we have applied a covariant formalism~\cite{Fuchs:2003qc,Pascalutsa:2006up,Geng:2008mf,Geng:2009hh} to successfully improve the classical Coleman-Glashow description of the baryon-octet magnetic moments by including the leading SU(3)-breaking provided by the chiral loops without~\cite{Geng:2008mf} and with explicit decuplet-baryon contributions~\cite{Geng:2009hh}.  This approach that includes both octet and decuplet virtual contributions has also been used to predict, up to $\mathcal{O}(p^4)$, the vector hyperon decay charge $f_1(0)$~\cite{Geng:2009ik}, which is essential to extract the Cabibbo-Kobayashi-Maskawa matrix element $V_{us}$ from the hyperon decay data.

The goal of the present paper is to use the covariant $\chi$PT formalism to describe the leading SU(3)-breaking (up to $\mathcal{O}(p^3)$) of the electromagnetic static properties of the decuplet baryons, and more particularly, of the $\Delta(1232)$ resonances. In Section II we display the chiral Lagrangians used in this work, discuss the power-counting problems and solutions of the present covariant calculation and introduce the electromagnetic form factors and moments of a spin-3/2 particle. In Section III we present the details of the calculation and the results for the MDMs, the EQMs, the MOMs and the CRs. The latter can be numerically achieved only after fixing the different low-energy constants (LECs) appearing up to this order. The single LEC that contributes to the MDMs will be fixed with the well measured $\mu_{\Omega^-}$, whereas the ones that contribute to the EQMs and to the CRs could be determined using lQCD results for the $\Omega^-$ at the physical point. Finally, there is no exclusive contribution of any LEC to the decuplet MOMs and they come as a true prediction of $\chi$PT at $\mathcal{O}(p^3)$.

\section{Formalism}
\subsection{Chiral Lagrangians}

The baryon-decuplet consists of a SU(3)-flavor multiplet of spin-3/2 resonances that we will represent with the Rarita-Schwinger field $T_\mu\equiv T^{ade}_\mu$ with the following associations:
$T^{111}=\Delta^{++}$, $T^{112}=\Delta^+/\sqrt{3}$,
$T^{122}=\Delta^0/\sqrt{3}$, $T^{222}=\Delta^-$, $T^{113}=\Sigma^{*+}/\sqrt{3}$,
$T^{123}=\Sigma^{*0}/\sqrt{6}$, $T^{223}=\Sigma^{*-}/\sqrt{3}$, 
$T^{133}=\Xi^{*0}/\sqrt{3}$, $T^{233}=\Xi^{*-}/\sqrt{3}$, and $T^{333}=\Omega^-$. The covariantized free Lagrangian is
\begin{equation}
 \mathcal{L}_{D}=\bar{T}^{abc}_\mu(i\gamma^{\mu\nu\alpha}D_\alpha-M_D\gamma^{\mu\nu})T^{abc}_\nu, \label{Eq:RSLag}
\end{equation}
where $M_D$ is decuplet-baryon mass and $D_\nu T_\mu^{abc}=\partial_\nu T_\mu^{abc}+(\Gamma_\nu,T_\mu)^{abc}$, $\Gamma_\nu$ being the chiral connection (see e.g. Ref.~\cite{Scherer:2002tk}) and with the definition $(X,T_\mu)^{abc}\equiv(X)_d^a T_\mu^{dbc}
+(X)_d^b T_\mu^{adc}+(X)_d^c T_\mu^{abd}$. In the last and following Lagrangians we sum any repeated SU(3)-index denoted by Latin characters $a,b,c,\ldots$, and $(X)^a_b$ denotes the element of the row $a$ and column $b$ of the matrix representation of $X$. 

For the meson-octet-decuplet and meson-decuplet-decuplet vertices we use the ``consistent'' lowest-order couplings~\cite{Pascalutsa:1999zz,Pascalutsa:2000kd,Pascalutsa:2006up}
\begin{eqnarray}
&&\mathcal{L}^{(1)}_{\phi B D}=\frac{i\,\mathcal{C}}{M_D F_\phi}\;\varepsilon^{abc}\left(\partial_\alpha\bar{T}^{ade}_\mu\right)\gamma^{\alpha\mu\nu}
 B^e_c\,\partial_\nu\phi^d_b+{\rm h.c.},\label{Eq:MBDCnsLag}\\
&&\mathcal{L}^{(1)}_{\phi D D}=\frac{i\mathcal{H}}{M_D F_\phi}\bar{T}_\mu^{abc}\gamma^{\mu\nu\rho\sigma}\gamma_5\left(\partial_\rho T_\nu^{abd}\right)\partial_\sigma\phi^c_d, \label{Eq:MDDCnsLag}
\end{eqnarray}
with $\phi$ and $B$ the SU(3) matrix representation of the pseudoscalar mesons and of the octet-baryons respectively and where $\mathcal{C}$ and $\mathcal{H}$ are the $\phi B D$ and $\phi D D$ couplings and $F_\phi$ is the meson-decay constant. Up to third order there are three terms that contribute to the observables studied in this paper
\begin{eqnarray}
&&\mathcal{L}_{\gamma D D}^{(2)}=-\frac{g_{d}}{8 M_D}  \bar{T}^{abc}_\mu\sigma^{\rho\sigma}g^{\mu\nu} (F_{\rho\sigma}^+, T_\nu)^{abc},\label{Eq:ChLag2nd}\\
&&\mathcal{L}_{\gamma D D}^{(3)}=-\frac{g_q}{16M_D^2} \bar{T}^{abc}_\mu \gamma^{\mu\rho\sigma}\left((\partial^\nu F_{\rho\sigma}^+), T_\nu\right)^{abc} -\frac{g_{er}}{12}\bar{T}^{abc}_\mu \gamma^{\mu\nu\sigma}\left((\partial^\rho F_{\rho\sigma}^+), T_\nu\right)^{abc}, \label{Eq:ChLag3rd}
\end{eqnarray}
with $F_{\mu\nu}^+=2eQ_qF_{\mu\nu}$, $e$ the fundamental electric charge, $Q_q$ the SU(3)-flavor quark-charge matrix and $F_{\mu\nu}$ the electromagnetic tensor. The LEC $g_d$ gives at $\mathcal{O}(p^2)$ the SU(3)-symmetric description of the anomalous part of the MDMs of the decuplet baryons, while the LECs $g_q$ and $g_{er}$ appear at $\mathcal{O}(p^3)$ and describe a SU(3)-symmetric part of the EQMs and CRs respectively. Up to $\mathcal{O}(p^3)$ there is not any unknown contact interaction (LEC) contributing exclusively to the MOM and, therefore it comes as a prediction from the chiral loops obtained in the present work. Finally, it is worth to observe that working out the flavor-index summations in Eqs. (\ref{Eq:ChLag2nd}) and (\ref{Eq:ChLag3rd}), we find that the SU(3)-symmetric contribution to the observables is proportional to the charge of the particular decuplet-baryon (see e.g. Ref.~\cite{Butler:1993ej}). 

The $\phi B D$ coupling is obtained by fitting the $\Delta\rightarrow N\pi$ decay
width~\cite{Geng:2009hh} which yields $\mathcal{C}\approx1.0$. The $\phi D D$ coupling $\mathcal{H}$ is barely known and we fix it using the large $N_c$ relation between the nucleon and $\Delta$ axial charges, $g_A$ and $H_A$ respectively, $H_{A}=(9/5)g_{A}$. Given that $H_A=2\mathcal{H}$ and $g_{A}=1.26$, we use $\mathcal{H}=1.13$. For the meson decay constants we take an average $F_\phi\equiv1.17f_\pi$ with $f_\pi=92.4$ MeV. For the masses of the pseudoscalar mesons we take
$m_\pi\equiv m_{\pi^\pm}=0.13957$ GeV, $m_K\equiv m_{K^\pm}=0.49368$ GeV, 
$m_\eta=0.5475$ GeV while for the baryon masses we use the average among the members of the respective SU(3)-multiplets, $M_B=1.151$ GeV and $M_D=1.382$ GeV.

\subsection{Power Counting}

We apply the standard power counting where one assigns a chiral order $n_{\chi PT}=4L-2N_M-N_B+\sum_k k V_k$ to a diagram with $L$ loops, $N_M$ ($N_B$) internal meson (octet- and decuplet-baryon) propagators and $V_k$ vertices from $k$th order Lagrangians. In the covariant theory with the modified minimal subtraction method ($\overline{MS}$), this rule is violated by lower-order analytical pieces~\cite{Gasser:1987rb}. In order to recover the power counting, we absorb into the LECs the terms breaking the power counting that are obtained expanding the loop-functions around the chiral limit (all the SU(3)-symmetric contribution of the loops)~\cite{Geng:2009hh} in a dimensional-regularization scheme known as the extended-on-mass-shell (EOMS) prescription~\cite{Fuchs:2003qc}. The regularized loops will then start to contribute at the order assigned by the power-counting but will also include higher-order corrections required by relativity and analyticity. We notice that only for the MDMs a power-counting restoration procedure is necessary since it is the only observable for which $\mathcal{O}(p^2)$ analytical chiral pieces (LECs) are possible.
 
Besides, the propagator corresponding to the RS action in $d$ dimensions
\begin{equation}
S^{\mu\nu}(p)=-\frac{\slashp+M_D}{p^2-M_D^2+i\epsilon}\left[
g^{\mu\nu}-\frac{1}{d-1}\gamma^\mu\gamma^\nu-\frac{1}{(d-1)M_D}\left(\gamma^\mu\, p^\nu-\gamma^\nu\, p^\mu\right)
-\frac{d-2}{(d-1)M_D^2} p^\mu p^\nu \right],
\end{equation}
has a problematic high-energy behavior. In the context of an effective field theory, this is responsible for the appearance of $d$ - 4 singularities of a chiral order higher than the one naively expected using the power counting rule explained above. These infinities would be absorbed by the proper higher-order counter-terms. However, we do not include these terms explicitly but perform a $\overline{MS}$-subtraction on them and study the uncertainty brought by the residual regularization-scale dependence.

\subsection{Spin-3/2 electromagnetic form factors}

\begin{figure}[t]
\includegraphics[width=\columnwidth]{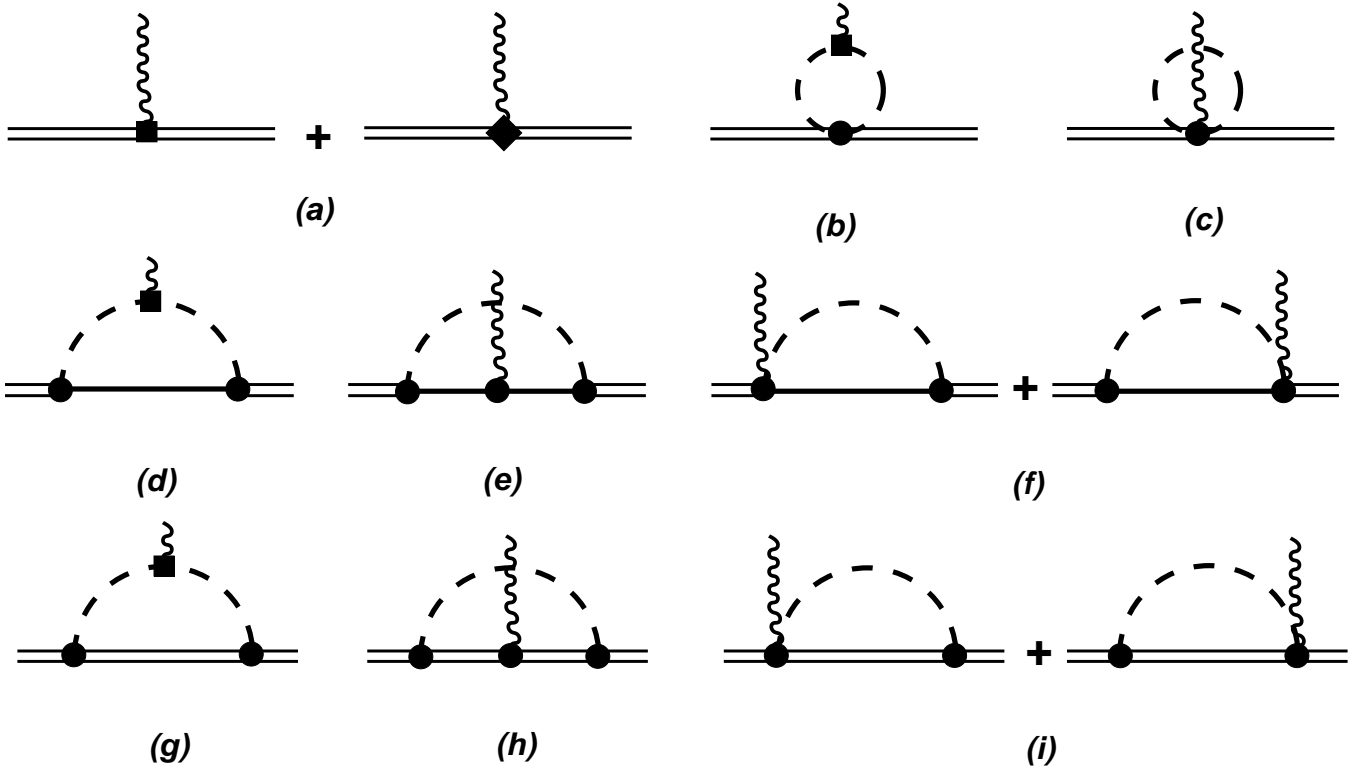}
\caption{Feynman diagrams that contribute up to $\mathcal{O}(p^3)$ to the decuplet electromagnetic form factors. The external double solid lines correspond to decuplet baryons, whereas the internal single (double) solid lines correspond to octet (decuplet) baryons. The dashed lines represent mesons. Black circles, black squares and black diamonds represent first-, second- and third-order couplings respectively. 
\label{fig:diagram}}
\end{figure}

The structure of the spin-3/2 particles, as probed by photons, is encoded into four electromagnetic form factors~\cite{Nozawa:1990gt}:
\begin{equation}
\langle T(p')|J^\mu|T(p)\rangle=-\bar{u}_\alpha(p')\Big\{\Big[F_1^*(\tau)\gamma^\mu+\frac{i\sigma^{\mu\nu}q_\nu}{2M_D}F_2^*(\tau)\Big]g^{\alpha\beta}+\Big[F_3^*(\tau)\gamma^\mu+\frac{i\sigma^{\mu\nu}q_\nu}{2M_D}F_4^*(\tau)\Big]\frac{q^\alpha q^\beta}{4M_D^2}\Big\}u_\beta(p),  \label{Eq:DeltaFF0}
\end{equation}
where $u_\alpha$ are the Rarita-Schwinger spinors and $\tau=-q^2/(4M_D^2)$. We can define the electric monopole and quadrupole and the magnetic dipole and octupole form factors in terms of the $F_i^*$'s:
\begin{eqnarray}
&&G_{E0}(\tau)=(F_1^*(\tau)-\tau F_2^*(\tau))+\frac{2}{3}\tau G_{E2}(\tau),\label{Eq:GEO}\\
&&G_{E2}(\tau)=(F_1^*(\tau)-\tau F_2^*(\tau))-\frac{1}{2}(1+\tau)(F_3^*(\tau)-\tau F_4^*(\tau)),\label{Eq:GE2}\\
&&G_{M1}(\tau)=(F_1^*(\tau)+F_2^*(\tau))+\frac{4}{5}\tau G_{M3}(\tau), \label{Eq:GM1}\\
&&G_{M3}(\tau)=(F_1^*(\tau)+F_2^*(\tau))-\frac{1}{2}(1+\tau)(F_3^*(\tau)+F_4^*(\tau)).  \label{Eq:GM3}
\end{eqnarray}
At $q^2=0$, the multipole form factors define the static electromagnetic moments, namely, the charge $Q$, the magnetic dipole moment $\mu$, the electric quadrupole moment $\mathcal{Q}$ and the magnetic octupole moment $O$
\begin{eqnarray}
&&Q=G_{E0}(0)=F_1^*(0),  \label{Eq:Dcharge}\\
&&\mu=\frac{e}{2M_D}G_{M1}(0)= \frac{e}{2M_D}(Q+F_2^*(0)), \label{Eq:DMDM}\\
&&\mathcal{Q}=\frac{e}{M_D^2}G_{E2}(0)=\frac{e}{M_D^2}(Q-\frac{1}{2}F_3^*(0)),\label{Eq:DQEM}\\
&&O=\frac{e}{2M_D^3}G_{M3}(0)=\frac{e}{2M_D^3}\left(G_{M1}(0)-\frac{1}{2}(F_3^*(0)+F_4^*(0))\right).\label{Eq:DOMM}
\end{eqnarray}

The electromagnetic multipole moments of the spin-3/2 resonances are connected with their spatial electromagnetic distributions and, therefore, with their internal structure. Particularly, the EQM and MOM measure the departure from a spherical shape of the charge and from a dipole magnetic distribution respectively.  

Besides the static electromagnetic moments, the slope of the form factors at $q^2=0$ is also of phenomenological interest. In particular the one corresponding to $G_{E0}$ is the so-called squared CR:
\begin{equation}
\langle r_{E0}^2\rangle=6\frac{dG_{E0}(q^2)}{dq^2}\Big|_{q^2=0}=6\frac{dF^*_{1}(q^2)}{dq^2}\Big|_{q^2=0}+\frac{3}{2M_D^2}F^*_2(0)-\frac{1}{M_D^2}G_{E2}(0).\label{Eq:ChRadiusDef}
\end{equation}

\section{Results}

The Feynman diagrams that give contribution to decuplet electromagnetic form factors are shown in Fig.~\ref{fig:diagram}. The loop contributions to any of the four form factors for a particular decuplet-baryon $D$, $\delta F^*_{j,D}(\tau)$ with $j=1,\ldots,4$ can be expressed as
\begin{eqnarray}
&&\delta F^*_{j,D}(\tau)=\frac{1}{\left(4 \pi F_\phi\right)^2 }\sum_{M=\pi,K}\xi^{(b)}_{DM}\left(H_j^{(b)}(\tau,m_M)+H_j^{(c)}(\tau,m_M)\right)+\nonumber\\
&&\frac{M_D^2}{\left(4 \pi F_\phi\right)^2}\left[\mathcal{C}^2\left(\sum_{M=\pi,K}\xi^{(d)}_{DM} H_j^{(d)}(\tau,m_M)+\sum_{M=\pi,K,\eta}\xi^{(e)}_{DM} \left(H_j^{(e)}(\tau,m_M)+H_j^{(f,II)}(\tau,m_M)\right)\right)+\right.\nonumber\\
&&\left.\mathcal{H}^2\left(\sum_{M=\pi,K}\xi^{(g)}_{DM} \left(H_j^{(g)}(\tau,m_M)+H_j^{(i,I)}(\tau,m_M)\right)+\sum_{M=\pi,K,\eta}\xi^{(h)}_{DM} \left(H_j^{(h)}(\tau,m_M)+H_j^{(i,II)}(\tau,m_M)\right)\right)\right],\nonumber\\
\label{Eq:ResGen}
\end{eqnarray}
with $H_j^{(X)}(\tau,m_M)$ the loop-function coming from the diagram $(X)$ and where the additional character that appears in the function of the diagrams \textbf{\textit{(f)}} and \textbf{\textit{(i)}} indicates whether the seagull-diagram comes from the minimal substitution performed on the derivative of the meson fields ($I$) or of the decuplet fields ($II$). The loop-functions for $j=2,\ldots,4$ at $q^2=0$, the first derivative with respect to $q^2$ of the one for $j=1$ at $q^2=0$ and the corresponding coefficients $\xi^{(X)}_{DM}$ are given in the Appendix in Table \ref{Table:Coefficients}. All this information together with Eq. (\ref{Eq:ResGen}), is what is required for obtaining, through the Eqs. (\ref{Eq:GE2})-(\ref{Eq:ChRadiusDef}), the loop results of the observables discussed in the present work. The contact interactions, diagrams \textbf{\textit{(a)}} in Fig.~\ref{fig:diagram}, provide the SU(3)-symmetric contribution and ensure the regularization of the divergences coming from the loops up to $\mathcal{O}(p^3)$. As explained above, they also allow to recover the power-counting by applying a suitable regularization prescription.

We have done some checks on the calculation of the loops of Fig.~\ref{fig:diagram}. The first one concerns the electromagnetic gauge invariance as well as the completeness of the Lorentz decomposition of Eq. (\ref{Eq:DeltaFF0}). Besides the structures collected there, one also obtains contributions to $g^{\alpha\mu}q^\beta$ and $g^{\beta\mu}q^\alpha$ and to the electromagnetic-gauge violating ones, $g^{\alpha\beta}q^\mu$ and $q^\alpha q^\beta q^\mu$. In order to fit the results of the loops into the representation (\ref{Eq:DeltaFF0}) we have used that~\cite{Nozawa:1990gt}
\begin{equation}
 g^{\beta\mu}q^\alpha=g^{\alpha\mu}q^\beta+2M_D(1+\tau)g^{\alpha\beta}\gamma^\mu-g^{\alpha\beta}P^\mu+\frac{1}{M_D}\gamma^\mu q^\alpha q^\beta \label{Eq:unnId}
\end{equation}
where $P^\mu=p^\mu+p'^\mu$ and obtained that the resulting coefficients of $g^{\alpha\mu}q^\beta$, $g^{\alpha\beta}q^\mu$ and $q^\alpha q^\beta q^\mu$ are identically zero. On top of that, we have tested the electromagnetic-gauge invariance by checking that the loop contributions to the electric charge vanish after including the wave-function renormalization $\Sigma'_D$. Indeed, for each decuplet-baryon $D$ of electric charge $Q_D$, we get $\delta F_{1,D}^*(0)+Q_D\Sigma'_D=0$. 

We have obtained that the following relations, which are a consequence of the assumed isospin symmetry, are fulfilled for any of the observables $\mathcal{X}$ studied in this work
\begin{eqnarray}
&&\mathcal{X}_{\Delta^{++}}- \mathcal{X}_{\Delta^{+}}-\mathcal{X}_{\Delta^{0}}+\mathcal{X}_{\Delta^{-}}=0,\nonumber\\
&&\mathcal{X}_{\Delta^{++}}- \mathcal{X}_{\Delta^{-}}-3(\mathcal{X}_{\Delta^{+}}-\mathcal{X}_{\Delta^{0}})=0,\nonumber\\
&&2\mathcal{X}_{\Sigma^{*0}}=\mathcal{X}_{\Sigma^{*+}}+\mathcal{X}_{\Sigma^{*-}}.\label{Eq:IsospRel}
\end{eqnarray}
Furthermore, among the SU(3)-flavor relations discussed in Ref.~\cite{Banerjee:1995wz} only two  
\begin{eqnarray}
&&\mathcal{X}_{\Delta^{0}}+\mathcal{X}_{\Xi^{*0}}=0, \label{Eq:SU3Rel1}\\
&&\mathcal{X}_{\Sigma^{*0}}=0, \label{Eq:Sigma0}
\end{eqnarray}
still hold when the higher-order relativistic corrections are incorporated. The Eqs.~(\ref{Eq:IsospRel})-(\ref{Eq:Sigma0}) mean that only the form factors of five of the ten decuplet resonances are really independent in $\mathcal{O}(p^3)$ 
covariant $\chi$PT. For the $\mathcal{O}(p^3)$ heavy-baryon expansion only two of them are independent~\cite{Banerjee:1995wz}.

The numerical results that we present in the following are obtained fixing the renormalization scale at $\mu=1$ GeV and using the values for the different masses and couplings displayed above. In the presented results, we also include an uncertainty estimated varying the renormalization scale and the mean baryon mass (keeping the mass splitting $M_D-M_B=0.231$ GeV fixed) in the intervals 0.7 GeV$\leq\mu\leq$1.3 GeV and 1 GeV$\leq M_B\leq$1.3 GeV. 

\begin{table*}
\centering
\caption{Values in nuclear magnetons ($\mu_N$) of the different contributions to the magnetic dipole moments of $\Delta^{++}$, $\Delta^+$, $\Sigma^{*+}$, $\Xi^{*-}$ and $\Omega^-$ after fitting the value of $\hat{g}_d$ to obtain $\mu_{\Omega^-}=-2.02(5)$. For the MDM of each baryon we show the results either in heavy-baryon or covariant $\chi$PT separated into the $\mathcal{O}(p^2)$ tree-level (TL) contribution, the $\mathcal{O}(p^3)$ chiral loop contributions coming from internal octet-baryons (O) and the $\mathcal{O}(p^3)$ chiral loop contributions coming from internal decuplet-baryons (D). We also list the fitted value of $\hat{g}_d$. \label{Table:ResMDMChPT}} 
\begin{ruledtabular}
\begin{tabular}{cc|ccc|ccc|ccc|ccc|ccc}
&&\multicolumn{3}{c}{$\Delta^{++}$}&\multicolumn{3}{c}{$\Delta^+$}&\multicolumn{3}{c}{$\Sigma^{*+}$}&\multicolumn{3}{c}{$\Xi^{*-}$}&\multicolumn{3}{c}{$\Omega^-$}\\
 \hline
 &$\hat{g}_d$&TL&O&D&TL&O&D&TL&O&D&TL&O&D&TL&O&D\\
\hline
\multicolumn{1}{c}{HB$\chi$PT $\mathcal{O}(p^3)$}&7.64&$11.75$&$-2.85$&$-0.96$&$5.87$&$-1.98$&$-0.57$&$5.87$&$-0.86$&$-0.39$&$-5.87$&$+1.98$&$+0.57$&$-5.87$&$+3.11$&$+0.75$\\
\multicolumn{1}{c}{Cov. $\chi$PT $\mathcal{O}(p^3)$}&4.71&$7.76$&$-1.09$&$-0.63$&$3.88$&$-0.70$&$-0.34$&$3.88$&$-0.46$&$-0.35$&$-3.88$&$+0.89$&$+0.44$&$-3.88$&$+1.34$&$+0.52$\\
\end{tabular}
\end{ruledtabular}
\end{table*}

\begin{table*}
\centering
\caption{Values in nuclear magnetons ($\mu_N$) of the decuplet magnetic dipole moments in relativistic chiral perturbation theory up to $\mathcal{O}(p^3)$ calculated in this work. We compare our results with the SU(3)-symmetric description and with those obtained in other theoretical approaches including the NQM~\cite{Hikasa:1992je}, the RQM~\cite{Schlumpf:1993rm}, the $\chi$QM ~\cite{Wagner:2000ii}, the $\chi$QSM~\cite{Ledwig:2008es}, the QCD-SR~\cite{Lee:1997jk}, (extrapolated) lQCD~\cite{Leinweber:1992hy,Lee:2005ds}, large $N_c$~\cite{Luty:1994ub} and the HB$\chi$PT calculation of Ref.~\cite{Butler:1993ej}. The experimental values are also included for reference~\cite{Yao:2006px}.  \label{Table:ResMDM}}
\begin{ruledtabular}
\begin{tabular}{ccccccccccc}
&$\Delta^{++}$&$\Delta^+$&$\Delta^0$&$\Delta^-$&$\Sigma^{*+}$&$\Sigma^{*0}$&$\Sigma^{*-}$&$\Xi^{*0}$&$\Xi^{*-}$&$\Omega^-$\\
\hline
\multicolumn{1}{c}{SU(3)-symm.}&4.04&2.02&0&-2.02&2.02&0&-2.02&0&-2.02&-2.02\\
\multicolumn{1}{c}{NQM~\cite{Hikasa:1992je}}&5.56&2.73&-0.09&-2.92&3.09&0.27&-2.56&0.63&-2.2&-1.84\\
\multicolumn{1}{c}{RQM~\cite{Schlumpf:1993rm}}&4.76&2.38&0&-2.38&1.82&-0.27&-2.36&-0.60&-2.41&-2.35\\
\multicolumn{1}{c}{$\chi$QM~\cite{Wagner:2000ii}}&6.93&3.47&0&-3.47&4.12&0.53&-3.06&1.10&-2.61&-2.13\\
\multicolumn{1}{c}{$\chi$QSM~\cite{Ledwig:2008es}} &4.85&2.35&-0.14&-2.63&2.47&-0.02&-2.52&0.09&-2.40&-2.29\\
\multicolumn{1}{c}{QCD-SR~\cite{Lee:1997jk}}&4.1(1.3)&2.07(65)&0&-2.07(65)&2.13(82)&-0.32(15)&-1.66(73)&-0.69(29)&-1.51(52)&-1.49(45)\\
\multicolumn{1}{c}{lQCD~\cite{Leinweber:1992hy}}&6.09(88)&3.05(44)&0&-3.05(44)&3.16(40)&0.329(67)&-2.50(29)&0.58(10)&-2.08(24)&-1.73(22)\\
\multicolumn{1}{c}{lQCD~\cite{Lee:2005ds}}&5.24(18)&0.97(8)&-0.035(2)&-2.98(19)&1.27(6)&0.33(5)&-1.88(4)&0.16(4)&-0.62(1)&---\\
\multicolumn{1}{c}{large $N_c$~\cite{Luty:1994ub}}&5.9(4)&2.9(2)&---&-2.9(2)&3.3(2)&0.3(1)&-2.8(3)&0.65(20)&-2.30(15)&-1.94\\
\multicolumn{1}{c}{HB$\chi$PT~\cite{Butler:1993ej}}&4.0(4)&2.1(2)&-0.17(4)&-2.25(19)&2.0(2)&-0.07(2)&-2.2(2)&0.10(4)&-2.0(2)&-1.94\\
\hline
\multicolumn{1}{c}{This work}&6.04(13)&2.84(2)&-0.36(9)&-3.56(20)&3.07(12)&0&-3.07(12)&0.36(9)&-2.56(6)&-2.02\\
\multicolumn{1}{c}{Expt.~\cite{Yao:2006px}}&5.6$\pm$1.9&$2.7^{+1.0}_{-1.3}\pm1.5\pm3$&---&---&---&---&---&---&---&-2.02$\pm0.05$\\
\end{tabular}
\end{ruledtabular}
\end{table*}

\subsection{Magnetic dipole moments}

The MDMs are the only observable discussed in this work for which there exist experimental data. More precisely, the MDM of the $\Delta^{++}$, the $\Delta^{+}$ and the $\Omega^-$ have been measured.  In order to obtain the MDMs of the different members of the decuplet in covariant $\chi$PT, we calculate the contributions to $F^*_2(0)$ of the diagrams listed in Figure \ref{fig:diagram} and use Eq. (\ref{Eq:DMDM}). Since we have a contact-term contribution at $\mathcal{O}(p^2)$ through the LEC $g_d$ whereas the loops start to contribute at $\mathcal{O}(p^3)$, we apply the power-counting restoration prescription explained in section II-B. After removing the $\mathcal{O}(p^2)$ ultraviolet divergences by the $\overline{MS}$ procedure, this is equivalent to redefine $g_d$ as
\begin{equation}
\hat{g}_d=g_d+\frac{\mathcal{C}^2M_D^2}{(4 \pi F_\phi)^2}f_d^1(\mu)+\frac{\mathcal{H}^2M_D^2}{(4 \pi F_\phi)^2}f_d^2(\mu)\label{Eq:RenMDM}
\end{equation}
where the definition of the functions $f^i_d(\mu)$ can be found in the Appendix. From the renormalized loop functions $\hat{H}^{(X)}$ we can then obtain the heavy-baryon expressions applying that $M_D=M_B+\delta$ and $M_B\sim\Lambda_{\chi SM}$ in what nowadays is called the small-scale expansion (SSE)~\cite{Hemmert:1997ye}. Only the diagrams \textbf{\textit{(d)}} and \textbf{\textit{(g)}} contribute up to $\mathcal{O}(p^3)$
\begin{eqnarray}
&&\hat{H}^{(d)}(m)\simeq \bar{\delta}\, r\log\left(\frac{\mu_m^2}{4\bar{\delta}^2}\right)+\left\{\begin{array}{c}
                             2\,r\,\sqrt{\mu_m^2-\bar{\delta}^2}\left(\frac{\pi}{2}+\arctan\left(\frac{\bar{\delta}}{\sqrt{\mu_m^2-\bar{\delta}^2}}\right)\right)\hspace{1.3cm}m\geq\delta \\
r\sqrt{\bar{\delta}^2-\mu_m^2}\left(-2\pi i+\log\left(\frac{\bar{\delta}+\sqrt{\bar{\delta}^2-\mu_m^2}}{\bar{\delta}-\sqrt{\bar{\delta}^2-\mu_m^2}}\right)\right)\hspace{1.3cm}m<\delta
                             \end{array}\right.,\label{Eq:HBexpansionB}\\
&&\hat{H}^{(g)}(m)\simeq\frac{2\,r\,\pi \,\mu_m}{3},\label{Eq:HBexpansionD}
\end{eqnarray}
where $r=M_B/M_D$, $\bar{\delta}=\delta/M_D$ and $\mu_m=m/M_D$. These loop-functions are equal to the ones found in~\cite{Banerjee:1995wz} and, with the coefficients $\xi^{(X)}_{DM}$ of Table \ref{Table:Coefficients}, they lead to the HB$\chi$PT results given in Table~\ref{Table:ResMDMChPT}\footnote{We have found discrepancies among the relative signs and absolute factors of the dynamical-octet and -decuplet diagrams reported in previous works~\cite{Butler:1993ej,Banerjee:1995wz,Tiburzi:2009yd,Cloet:2003jm}. Besides, the loop function coming from \textit{\textbf{(d)}} in Fig.~\ref{fig:diagram} is multivalued and we have noticed that the form presented in Refs.~\cite{Butler:1993ej,Tiburzi:2009yd} does not univocally give the physical branches. These are specified in Eq.~(\ref{Eq:HBexpansionB}). Notice that the loop functions develop an imaginary part for $m<\delta$, although in the present work we only discuss the real part.}. Since the only precise experimental value on the decuplet MDMs is used to determine the unknown LEC  $\hat{g}_d$, it is not really possible to directly compare the quality of the HB$\chi$PT and covariant $\chi$PT results confronted to experimental data. Nonetheless, we can compare the convergence properties of both schemes. In Table \ref{Table:ResMDMChPT} we list the results for the MDMs of $\Delta^{++}$, $\Delta^+$, $\Sigma^{*+}$, $\Xi^{*-}$ and $\Omega^-$ after fitting the value of $\hat{g}_d$ to obtain $\mu_{\Omega^-}=-2.02(5)$. The results for the rest of the members of the decuplet can be obtained using Eqs. (\ref{Eq:IsospRel})-(\ref{Eq:Sigma0}). For the MDM of any of the baryons we show the results either in HB$\chi$PT or covariant $\chi$PT separated into the $\mathcal{O}(p^2)$ tree-level (TL) contribution, the $\mathcal{O}(p^3)$ chiral correction coming from internal octet-baryons (O) and the $\mathcal{O}(p^3)$ chiral correction coming from internal decuplet-baryons (D). 

For any of the five baryons displayed in Table \ref{Table:ResMDMChPT}, we observe that the heavy-baryon loop contributions are larger than the covariant ones. The main difference arises from the loops with internal octet-baryons for which HB$\chi$PT gives more than two times the covariant approach for most of the channels. The chiral corrections with internal decuplet-baryons in the two schemes are rather more similar, with the HB$\chi$PT-SSE ones about 50$\%$ larger than those obtained in the covariant calculation. Particularly for the $\Delta^{++}$, we find that the heavy-baryon prediction $\mu_{\Delta^{++}}=7.94\,\mu_N$ is bigger than the upper bound provided by the PDG, $\mu_{\Delta^{++}}\le7.5\mu_N$~\cite{Yao:2006px}. These comparisons suggest that the heavy-baryon expansion probably overestimates the size of the chiral corrections to the MDMs of the decuplet resonances as it occured for the case of the baryon-octet magnetic moments~\cite{Geng:2008mf,Geng:2009hh}. The comparison with the heavy-baryon study of Ref.~\cite{Butler:1993ej} is not straightforward since in the latter the physical baryon masses as well as the physical meson-decay constants are used, which accounts for higher-order SU(3)-breaking mechanisms not included in the present work. The strict third-order HB$\chi$PT-SSE results are the ones presented in Table~\ref{Table:ResMDMChPT}.

In Table \ref{Table:ResMDM} we compare the results obtained in the covariant $\chi$PT approach of the present work for the MDMs of all the decuplet-baryons with the ones obtained in NQM~\cite{Hikasa:1992je}, RQM~\cite{Schlumpf:1993rm}, $\chi$QM ~\cite{Wagner:2000ii}, $\chi$QSM~\cite{Ledwig:2008es}, QCD-SR~\cite{Lee:1997jk}, (extrapolated~\footnote{It must be pointed out that the chiral extrapolations in Refs.~\cite{Leinweber:1992hy,Lee:2005ds} have been performed without taking into account the non-trivial analytical structure across the point $m=M_D-M_N$~\cite{Pascalutsa:2004je,Cloet:2003jm} and the artifacts introduced by the quenched approximation at such values of $m$ (see for instance~\cite{Boinepalli:2009sq}). Therefore their results should be compared with some care.}) quenched lQCD~\cite{Leinweber:1992hy,Lee:2005ds}, large $N_c$~\cite{Luty:1994ub} and the HB$\chi$PT calculation of Ref.~\cite{Butler:1993ej}. We also list the experimental values as averaged by the PDG~\cite{Yao:2006px} . In general, our results are consistent with the central value of the experimental numbers for $\mu_{\Delta^{++}}$ and $\mu_{\Delta^+}$. Moreover, for the former we do agree very satisfactorily with the latest experiment, $\mu_{\Delta^{++}}=6.14\pm0.51$~\cite{LopezCastro:2000cv}. The covariant $\chi$PT results are also consistent with those obtained in other approaches, although they tend to be larger for all channels. Interestingly, they are very similar to the ones obtained in the large $N_c$ expansion of Ref.~\cite{Luty:1994ub} and also to those reported in the NQM~\cite{Hikasa:1992je}. It is  worth to notice that the higher-order uncertainties of the covariant $\chi$PT results for the MDMs given by the chosen values for $M_B$, $M_D$ and $\mu$ are very small. 

The present work is also to be compared with studies focused on the MDM of the $\Delta(1232)$ resonance. We find  again that the values predicted in covariant $\chi$PT are larger than those found in lQCD ($\mu_{\Delta^+}=2.32(16)\mu_N$~\cite{Alexandrou:2008bn}, $\mu_{\Delta^+}=2.49(27)\mu_N$~\cite{Cloet:2003jm}), in the SpQM ($\mu_{\Delta^+}=2.51\mu_N$~\cite{Ramalho:2008dc}) and with light cone QCD-SRs ($\mu_{\Delta^+}=2.2(4)\mu_N$~\cite{Aliev:2000rc}).

\subsection{Electric quadrupole moments}

\begin{table*}
\centering
\caption{Values of the electric quadrupole moments of the decuplet resonances in relativistic chiral perturbation theory up to $\mathcal{O}(p^3)$ (in units of $10^{-2}$ fm$^2$). We express the results in terms of the quadrupole moment of the $\Omega^-$.  \label{Table:ResEQM}}
\begin{ruledtabular}
\begin{tabular}{ccccc}
$\Delta^{++}$&$\Delta^+$&$\Delta^0$&$\Delta^-$&$\Sigma^{*+}$\\
\hline
$-2 \mathcal{Q}_{\Omega^-}-0.9(3.3)$&$-\mathcal{Q}_{\Omega^-}-1.6(1.5)$&$-2.20(24)$&$\mathcal{Q}_{\Omega^-}-2.8(2.0)$&$-\mathcal{Q}_{\Omega^-}+1.9(1.3)$\\
\hline
\hline
$\Sigma^{*0}$&$\Sigma^{*-}$&$\Xi^{*0}$&$\Xi^{*-}$&$\Omega^-$\\
\hline
0&$\mathcal{Q}_{\Omega^-}-1.9(1.3)$&$2.20(24)$&$\mathcal{Q}_{\Omega^-}-1.0(0.6)$&$\mathcal{Q}_{\Omega^-}$\\
\end{tabular}
\end{ruledtabular}
\end{table*}

\begin{table*}
\centering
\caption{Values of the electric quadrupole moments in units of $10^{-2}$ fm$^2$ in different theoretical approaches. We compare the results obtained using the latest quenched lQCD (qlQCD) result~\cite{Boinepalli:2009sq} in combination with the relativistic chiral corrections (Table~\ref{Table:ResEQM}) with those obtained in the NQM~\cite{Krivoruchenko:1991pm}, in $\chi$QM ~\cite{Wagner:2000ii}, in GP~\cite{Buchmann:2002xq}, in light cone QCD-SR~\cite{Azizi:2008tx,Aliev:2009pd} and in HB$\chi$PT~\cite{Butler:1993ej}.\label{Table:ResEQMComp}} 
\begin{ruledtabular}
\begin{tabular}{ccccccccccc}
&$\Delta^{++}$&$\Delta^+$&$\Delta^0$&$\Delta^-$&$\Sigma^{*+}$&$\Sigma^{*0}$&$\Sigma^{*-}$&$\Xi^{*0}$&$\Xi^{*-}$&$\Omega^-$\\
\hline
NQM~\cite{Krivoruchenko:1991pm}&-9.3&-4.6&0&4.6&-5.4&-0.7&4.0&-1.3&3.4&2.8\\
$\chi$QM~\cite{Wagner:2000ii}&-25.2&-12.6&0&12.6&-12.3&-2.1&8.2&-3.0&4.8&2.6\\
GP~\cite{Buchmann:2002xq}&-22.6&-11.3&0&11.3&-10.7&-1.7&7.4&-2.3&4.4&2.4\\
QCD-SR~\cite{Azizi:2008tx,Aliev:2009pd}&-2.8(8)&-1.4(4)&0&1.4(4)&-2.5(8)&0.1(3)&3(1)&0.23(7)&4(1)&10(3)\\
HB$\chi$PT~\cite{Butler:1993ej}&-8(5)&-3(2)&1.2(5)&6(3)&-7(3)&-1.3(7)&4(2)&-3.5(2)&2(1)&0.9(5)\\
This work+qlQCD~\cite{Boinepalli:2009sq}&-2.7(3.3)&-2.4(1.5)&-2.20(24)&-2.0(2.0)&1.1(1.3)&0&-1.1(1.3)&2.20(24)&-0.1(6)&0.86\\
\end{tabular}
\end{ruledtabular}
\end{table*}

Although so far there is not experimental information on the EQMs of the decuplet, they have motivated several theoretical studies in the past. Their interest lie in that they provide information on the deviation from a spherical shape of the charge distribution  and, consequently, on the internal structure of the spin-3/2 resonances. To obtain the covariant $\chi$PT results for the EQMs it is required to determine the unknown LEC $g_q$ and use Eq.~(\ref{Eq:GE2}) after evaluating the loop contributions given by the diagrams of Fig.~\ref{fig:diagram}. The LEC $g_q$ could be fixed with an eventual experimental value of the EQM of one of the members of the decuplet-baryons, most likely the one of the $\Omega^-$ (for proposed experimental methods to measure it we refer to Ref.~\cite{Buchmann:2002xq} and references therein).  An alternative source of  information could come from lQCD since the properties of the $\Omega^-$ can be obtained at the physical point and, consequently, a full-dynamical lQCD (unquenched) calculation of its electromagnetic properties could be reached in the near future. Once this value is used to determine $g_q$, $\chi$PT provides a prediction on the EQMs of the rest of the decuplet-baryons and, in particular of the $\Delta(1232)$. Therefore, it is particularly interesting to express the $\chi$PT results of the EQMs for the decuplet in terms of the EQM of the $\Omega^-$. This can be done by just redefining $g_q$
\begin{equation}
 \hat{g}_q=g_q+\delta\mathcal{Q}_{\Omega^-} \label{Eq:EQMred}
\end{equation}
where $\delta\mathcal{Q}_{\Omega^-}$ is the loop contribution to the EQM of the $\Omega^-$, and $\hat{g}_q$ would then mean the physical $\mathcal{Q}_{\Omega^-}$.

In Table ~\ref{Table:ResEQM} we list the results obtained for the EQMs of the decuplet in relativistic $\chi$PT up to $\mathcal{O}(p^3)$. They consist of the SU(3)-symmetric part depending on the value $\mathcal{Q}_{\Omega^-}$ that we encourage to fix in the future using either experiment or unquenched lQCD, in addition to the leading relativistic loop contributions. If in a first approximation we use the recent quenched lQCD result $\mathcal{Q}_{\Omega^-}=0.86(12)10^{-2}$ fm$^2$~\cite{Boinepalli:2009sq} to fix $g_q$, we obtain the results displayed in Table~\ref{Table:ResEQMComp} compared with those obtained in NQM~\cite{Krivoruchenko:1991pm}, $\chi$QM~\cite{Wagner:2000ii}, QCD-SR~\cite{Azizi:2008tx,Aliev:2009pd} and HB$\chi$PT~\cite{Butler:1993ej}. We observe that with this value of $\mathcal{Q}_{\Omega^-}$, the loop contributions are quite large and the EQMs of the decuplet-baryons are dominated by the chiral SU(3)-breaking corrections. 

We can also compare  with calculations focused on the $\Delta(1232)$ isospin multiplet. The result on the $\Delta^+$ given in Table~\ref{Table:ResEQMComp},  $\mathcal{Q}_{\Delta^+}=-2.5(1.5)10^{-2}$ fm$^2$, marginally agrees with recent theoretical determinations within the $\chi$QSM ($\mathcal{Q}_{\Delta^+}=-5.09$ $10^{-2}$ fm$^2$~\cite{Ledwig:2008es}) and the SpQM ($\mathcal{Q}_{\Delta^+}=-4.2$ $10^{-2}$ fm$^2$~\cite{Ramalho:2009vc}). 

\subsection{Magnetic Octupole Moments}
 
The MOMs of the decuplet baryons are experimentally unknown and only few theoretical predictions are available. Their interest also lie in that they contain information on the internal structure of the spin-3/2 baryons, more precisely on the current and spin distribution beyond the dipole form one given by the MDMs. From the $\chi$PT perspective, there are no LECs contributing exclusively to the MOMs up to $\mathcal{O}(p^3)$, although they depend on the ones that contribute to the MDMs, $g_d$, and to the EQMs, $g_q$ (see Eq.~(\ref{Eq:DOMM})). Once these LECs are fixed, the MOMs come as a true prediction from the chiral loops in the covariant formalism. In the heavy-baryon scheme the loop contributions to the MOMs are at least of order $\mathcal{O}(p^4)$ so that the relativistic results could be considered from that perspective as pure recoil corrections. In Table ~\ref{Table:ResMOM} we show the results for the MOMs once $g_d$ is fixed with the $\Omega^-$ MDM and the $g_q$ dependence is introduced in terms of the $\Omega^-$ EQM (in the proper units $\tilde{\mathcal{Q}}=\left(\mathcal{Q}/M_D\right)\;\left[e/(2M_N^3)\right]$). If we use again the value obtained in quenched lQCD~\cite{Boinepalli:2009sq} for the $\Omega^-$ EQM, $\tilde{\mathcal{Q}}_{\Omega^-}=0.113\,e/(2M_N^3)$, we obtain the results displayed in the last row of Table~\ref{Table:ResMOMcomp}. Moreover, in the same table we also collect the ones obtained previously in the general parameterization method~\cite{Buchmann:2008zza} and in light-cone QCD sum-rules~\cite{Azizi:2008tx,Aliev:2009pd}. Our results for the $\Delta^+$ favour a negative value for the MOM of the $\Delta^+$, in agreement with those obtained in the two latter approaches. Remarkably, our prediction for $O_{\Omega^{-}}$ agrees with the recent determination from the same quenched lQCD calculation used to fix $\tilde{\mathcal{Q}}$, $O_{\Omega^{-}}=0.2(1.2) e/(2M_N)^3$~\cite{Boinepalli:2009sq}. 

\begin{table*}
\centering
\caption{Values in units of $e/(2M_N^3)$ of the magnetic octupole moments of the members of the decuplet resonances in relativistic chiral perturbation theory up to $\mathcal{O}(p^3)$. The results depend on the $\Omega^-$ electric quadrupole moment given in proper units, $\tilde{\mathcal{Q}}=\left(\mathcal{Q}/M_D\right)\;\left[e/(2M_N^3)\right]$.  \label{Table:ResMOM}}
\begin{ruledtabular}
\begin{tabular}{ccccc}
$\Delta^{++}$&$\Delta^+$&$\Delta^0$&$\Delta^-$&$\Sigma^{*+}$\\
\hline
$-2 \tilde{\mathcal{Q}}_{\Omega^-}-1.6(4.2)$&$-\tilde{\mathcal{Q}}_{\Omega^-}-0.8(2.1)$&$0.026(16)$&$\tilde{\mathcal{Q}}_{\Omega^-}+0.8(2.1)$&$-\tilde{\mathcal{Q}}_{\Omega^-}-0.5(2.0)$\\
\hline\hline
$\Sigma^{*0}$&$\Sigma^{*-}$&$\Xi^{*0}$&$\Xi^{*-}$&$\Omega^-$\\
\hline
0&$\tilde{\mathcal{Q}}_{\Omega^-}+0.5(2.0)$&$-0.026(16)$&$\tilde{\mathcal{Q}}_{\Omega^-}+0.3(1.9)$&$\tilde{\mathcal{Q}}_{\Omega^-}+0(1.7)$\\
\end{tabular}
\end{ruledtabular}
\end{table*}

\begin{table*}
\centering
\caption{Values in units of $e/(2M_N^3)$ of the magnetic octupole moments of the members of the decuplet resonances in different theoretical approaches.  \label{Table:ResMOMcomp}}
\begin{ruledtabular}
\begin{tabular}{ccccccccccc}
&$\Delta^{++}$&$\Delta^+$&$\Delta^0$&$\Delta^-$&$\Sigma^{*+}$&$\Sigma^{*0}$&$\Sigma^{*-}$&$\Xi^{*0}$&$\Xi^{*-}$&$\Omega^-$\\
\hline
GP~\cite{Buchmann:2008zza}&-5.2&-2.6&0&2.6&-0.87&0.43&1.7&0.43&1.1&0.7\\
LCQCD SR~\cite{Azizi:2008tx,Aliev:2009pd}&-1.3(4)&-0.65(21)&0&0.65(21)&-2.6(9)&-0.11(2)&2.6(9)&-0.28(11)&2.2(9)&3.3(1.1)\\
This work+qlQCD~\cite{Boinepalli:2009sq}&-1.8(4.2)&-0.9(2.1)&0.026(16)&1.0(2.1)&-0.7(2.0)&0&0.7(2.0)&-0.026(16)&0.4(1.9)&0.2(1.8)\\
\end{tabular}
\end{ruledtabular}
\end{table*}

\begin{table*}
\centering
\caption{Values in units of fm$^2$ of the squared CR of the members of the decuplet resonances in relativistic chiral perturbation theory up to $\mathcal{O}(p^3)$. We express the results in terms of the corresponding squared CR of the $\Omega^-$.  \label{Table:ResCR}}
\begin{ruledtabular}
\begin{tabular}{ccccc}
$\Delta^{++}$&$\Delta^+$&$\Delta^0$&$\Delta^-$&$\Sigma^{*+}$\\
\hline
$-2r^2_{\Omega^-}+0.035(13)$&$-r^2_{\Omega^-}+0.021(6)$&$0.006(1)$&$r^2_{\Omega^-}-0.009(8)$&$-r^2_{\Omega^-}+0.008(6)$\\
\hline\hline
$\Sigma^{*0}$&$\Sigma^{*-}$&$\Xi^{*0}$&$\Xi^{*-}$&$\Omega^-$\\
\hline
0&$r^2_{\Omega^-}-0.008(6)$&$-0.006(1)$&$r^2_{\Omega^-}-0.005(3)$&$r^2_{\Omega^-}$\\
\end{tabular}
\end{ruledtabular}
\end{table*}

\subsection{Charge radii}

In Table~\ref{Table:ResCR} we show the results for the leading breaking corrections to the SU(3)-symmetric description of the quadratic CR of the decuplet baryons expressed in terms of the one of the $\Omega^-$.  This can be done using a redefinition of the LEC $g_{cr}$ equivalent to the one performed for the EQMs, Eq.~(\ref{Eq:EQMred}).
This LEC could be determined either from experiment or, in a model independent way, from lQCD. A remarkable feature of the chiral corrections to the squared CR is that they are quite small. Taking the value from quenched lQCD for the $\Omega^-$, $r^2_{\Omega^-}=-0.307(15)$ ~\cite{Boinepalli:2009sq}, we observe that the calculated chiral loops represent less than a 10$\%$ correction to the SU(3)-symmetric prediction. Therefore, we may anticipate that the description of the CR is dominated by short-range physics. Moreover, using the value from the lattice we can predict the CR of the rest of the decuplet baryons and, in particular, of the $\Delta$(1232) isospin multiplet. Indeed, we obtain for the $\Delta^+$ a quadratic radii $r^2_{E0}=0.328(16)$ fm$^2$ that we can compare with recent results obtained in the $\chi$QM ($r^2_{E0}=0.781$ fm$^2$~\cite{Buchmann:1996bd}), the $\chi$QSM ($r^2_{E0}=0.794$ fm$^2$~\cite{Ledwig:2008es}), the SpQM ($r^2_{E0}=0.325$ fm$^2$~\cite{Ramalho:2008dc}) and in lQCD ($r^2_{E0}=0.477(8)$ fm$^2$~\cite{Alexandrou:2008bn}). 

\section{Summary and Conclusions}

In this work we have studied the electromagnetic static properties of the lowest-lying decuplet of baryons in covariant $\chi$PT, with special attention given to the $\Delta(1232)$ isospin multiplet. The MDMs are of most relevance since they are the only diagonal electromagnetic observables for which there exist some experimental information. More precisely, the MDM of the $\Omega^-$ has been measured with a good precision, while the values for the MDMs of the $\Delta^{++}$ and $\Delta^+$ are not very accurate yet.
By fixing the only LEC appearing up to $\mathcal{O}(p^3)$ with the MDM of the $\Omega^-$ the covariant $\chi$PT prediction is that $\mu_\Delta^{++}=6.04(13)$ and $\mu_\Delta^+=2.84(2)$, which are very close to the central values of the current PDG~\cite{Yao:2006px}. Moreover, our agreement with the latest experimental value for the $\Delta^{++}=$ $\mu_\Delta^{++}=6.14\pm0.51$~\cite{LopezCastro:2000cv} is excellent. Nevertheless, the PDG averages are still afflicted with large uncertainties within which the results coming from any of the theoretical approaches collected in Table~\ref{Table:ResMDM} are consistent. Therefore the new and high precision data for the MDM of the $\Delta^+$ that is expected to come soon~\cite{Kotulla:2008zz} will be extremely valuable to assess the quality of the different theoretical predictions. Among these different approaches, the large $N_c$~\cite{Luty:1994ub} and the NQM~\cite{Hikasa:1992je} give results that are more consistent with the ones obtained in the present work. 

We have also studied the higher-order electromagnetic multipoles, the EQMs and the MOMs, and the CR. These properties that give insight into the spin-3/2 internal structure have been receiving increasing attention lately. Although experimental data is not available yet and it's doubtful it will be in a near future, the rapid development of lQCD could lead soon to model-independent results on these observables. In covariant $\chi$PT, the EQMs, the MOMs and the CRs depend on two unknown LECs that we have related with the CR and the EQM of the $\Omega^-$, which is the decuplet baryon for which reliable information is expected to come sooner. With the current results obtained in quenched lQCD, we predict for the $\Delta(1232)$ values of these observables that are consistent with other approaches. In particular we predict negative values for the EQM and MOM of the $\Delta^+$, and a squared CR that is almost half that of the proton. Finally, concerning the future of lQCD in the evaluation of the observables discussed in this work, we want to stress the non-trivial analytical structure across the point $m=M_D-M_B$ unveiled by different $\chi$PT studies. In this regard we want to highlight that the present calculation provides for the first time the covariant $\chi$PT $\mathcal{O}(p^3)$ results including the contributions of both dynamical octet- and decuplet-baryons that may be helpful to extrapolate the lQCD results to the physical point.

\section{Acknowledgments}

This work was partially supported by the  MEC grant  FIS2006-03438 and the European Community-Research Infrastructure
Integrating Activity Study of Strongly Interacting Matter (Hadron-Physics2, Grant Agreement 227431) under the Seventh Framework Programme of EU. L.S.G. acknowledges support from the MICINN in the Program 
``Juan de la Cierva''. J.M.C. acknowledges the same institution for a FPU grant. 

\section{Appendix}

\subsection{Loop-functions}

\begin{table*}
\centering
\caption{Coefficients of the loop-contribution Eq.~(\ref{Eq:ResGen}) for any of the decuplet-baryons $D$. \label{Table:Coefficients}}
\begin{ruledtabular}
\begin{tabular}{ccccccccccc}
&$\Delta^{++}$&$\Delta^+$&$\Delta^0$&$\Delta^-$&$\Sigma^{*+}$&$\Sigma^{*0}$&$\Sigma^{*-}$&$\Xi^{*0}$&$\Xi^{*-}$&$\Omega^-$\\
\hline
$\xi^{(b)}_{\pi,D}$&$\frac{3}{4}$&$\frac{1}{4}$&$-\frac{1}{4}$&$-\frac{3}{4}$&$\frac{1}{2}$&0&$-\frac{1}{2}$&$\frac{1}{4}$&$-\frac{1}{4}$&0\\
$\xi^{(b)}_{K,D}$&$\frac{3}{4}$&$\frac{1}{2}$&$\frac{1}{4}$&0&$\frac{1}{4}$&0&$-\frac{1}{4}$&$-\frac{1}{4}$&$-\frac{1}{2}$&$-\frac{3}{4}$\\
$\xi^{(d)}_{\pi,D}$&$-4$&$-\frac{4}{3}$&$\frac{4}{3}$&$4$&$-\frac{8}{3}$&0&$\frac{8}{3}$&$-\frac{4}{3}$&$\frac{4}{3}$&0\\
$\xi^{(d)}_{K,D}$&$-4$&$-\frac{8}{3}$&$-\frac{4}{3}$&0&$-\frac{4}{3}$&0&$\frac{4}{3}$&$\frac{4}{3}$&$\frac{8}{3}$&$4$\\
$\xi^{(e)}_{\pi,D}$&$4$&$\frac{8}{3}$&$\frac{4}{3}$&0&$\frac{2}{3}$&0&$-\frac{2}{3}$&$-\frac{4}{3}$&$-\frac{2}{3}$&0\\
$\xi^{(e)}_{K,D}$&$4$&$\frac{4}{3}$&$-\frac{4}{3}$&$-4$&$\frac{4}{3}$&0&$-\frac{4}{3}$&$\frac{4}{3}$&$-\frac{4}{3}$&$-4$\\
$\xi^{(e)}_{\eta,D}$&0&0&0&0&$2$&0&$-2$&0&$-2$&0\\
$\xi^{(g)}_{\pi,D}$&$-\frac{4}{3}$&$-\frac{4}{9}$&$\frac{4}{9}$&$\frac{4}{3}$&$-\frac{8}{9}$&0&$\frac{8}{9}$&$-\frac{4}{9}$&$\frac{4}{9}$&0\\
$\xi^{(g)}_{K,D}$&$-\frac{4}{3}$&$-\frac{8}{9}$&$-\frac{4}{9}$&0&$-\frac{4}{9}$&0&$\frac{4}{9}$&$\frac{4}{9}$&$\frac{8}{9}$&$\frac{4}{3}$\\
$\xi^{(h)}_{\pi,D}$&$\frac{16}{3}$&$\frac{26}{9}$&$\frac{4}{9}$&$-2$&$\frac{8}{9}$&0&$-\frac{8}{9}$&$-\frac{4}{9}$&$-\frac{2}{9}$&0\\
$\xi^{(h)}_{K,D}$&$\frac{4}{3}$&$\frac{4}{9}$&$-\frac{4}{9}$&$-\frac{4}{3}$&$\frac{28}{9}$&0&$-\frac{28}{9}$&$\frac{4}{9}$&$-\frac{28}{9}$&$-\frac{4}{3}$\\
$\xi^{(h)}_{\eta,D}$&$\frac{4}{3}$&$\frac{2}{3}$&0&$-\frac{2}{3}$&0&0&0&0&$-\frac{2}{3}$&$-\frac{8}{3}$\\
\end{tabular}
\end{ruledtabular}
\end{table*}

In the calculation of the loop diagrams, we have used the following
$d$-dimensional integrals in Minkowski space:
\begin{equation}
\int d^d k\frac{k^{\alpha_1}\ldots k^{\alpha_{2n}}}{(\mathcal{M}^2-k^2)^\lambda}
=i\pi^{d/2}\frac{\Gamma(\lambda-n+\varepsilon-2)}{2^n\Gamma(\lambda)}\frac{(-1)^n g^{\alpha_1
\ldots\alpha_{2n}}_s}{(\mathcal{M}^2)^{\lambda-n+\varepsilon-2}}
\end{equation}
with $g^{\alpha_1 \ldots\alpha_{2n}}_s=g^{\alpha_1\alpha_2}\ldots g^{\alpha_{2n-1}\alpha_{2n}}+\ldots$
symmetrical with respect to the permutation of any pair of indices (with $(2n-1)!!$
terms in the sum). We will present the divergent part of the loops as the contact piece $\lambda_\varepsilon=2/\varepsilon+\log{4\pi}-\gamma_E$, where $\varepsilon=4-d$ and $\gamma_E\simeq0.5772$ the Euler constant. 

We display below the loop functions $H_j^{(X)}$ and $H_1^{'(X)}\equiv\partial_{q^2}H_1^{(X)}|_{q^2=0}$ of the diagrams of Fig.~\ref{fig:diagram} that contribute to the observables studied in this work. These are written in a dimensionless form using $M_B=r M_D$, $\mu=\bar{\mu} M_D$, $\mathcal{M}_B^2=x m^2 + (1 - x) M_B^2 - x(1 - x) M_D^2$, $\mathcal{M}_D^2=x m^2 + (1 - x)^2 M_D^2$ and $\mathcal{M}^2_{B,D}=M_D^2\bar{\mathcal{M}}^2_{B,D}$. These loop functions are
\begin{equation*}
H_1^{'(b)}=\frac{2}{3}\left(\lambda_\epsilon-\frac{1}{2}\log\left(\frac{m^2}{\mu^2}\right)\right),
\end{equation*}
\begin{equation*}
H_2^{(d)}=\frac{1}{2}\int^1_0 dx\,x (r+x) (2 x-1) \left(\lambda _{\epsilon }+\log \left(\frac{\bar{\mathcal{M}}_B^2}{\bar{\mu} ^2}\right)\right),
\end{equation*}
\begin{equation*}
H_3^{(d)}= \frac{1}{3}\int^1_0 dx\,x^2\left(\frac{2 (x-1) (r+x) x}{\bar{\mathcal{M}}_B^2}+(3 r+4 x) \left(\lambda _{\epsilon }+\log \left(\frac{\bar{\mathcal{M}}_B^2}{\bar{\mu} ^2}\right)\right)\right),
\end{equation*}
\begin{equation*}
H_4^{(d)}= -\int^1_0 dx\,\frac{2 (x-1) x^3 (r+x)}{3 \bar{\mathcal{M}}_B^2},
\end{equation*}
\begin{equation*}
H_1^{'(d)}= -\frac{1}{24M_D^2}\int^1_0 dx\,x^2 \left((3 r+2 x) \left(\lambda _{\epsilon }+\log \left(\frac{\bar{\mathcal{M}}_B^2}{\bar{\mu} ^2}\right)\right)-\frac{2(x-1) x (r+x)}{
   \bar{\mathcal{M}}_B^2}\right),
\end{equation*}
\begin{equation*}
H_2^{(e)}=-\int^1_0 dx\,(x-1)^2 (r+x) \left(\lambda _{\epsilon }+\log \left(\frac{\bar{\mathcal{M}}_B^2}{\bar{\mu} ^2}\right)\right),
\end{equation*}
\begin{equation*}
H_3^{(e)}= -\frac{1}{3}\int^1_0 dx\,(x-1)^2\left(\frac{(r+x)^2 (x-1)}{\bar{\mathcal{M}}_B^2}+ \left(1-x+3 (r+1) \left(\lambda _{\epsilon
   }+\log \left(\frac{\bar{\mathcal{M}}_B^2}{\bar{\mu}^2}\right)\right)\right)\right),
\end{equation*}
\begin{equation*}
H_4^{(e)}= \int^1_0 dx\,\frac{2 (x-1)^4 (r+x)}{3\bar{\mathcal{M}}_B^2 },
\end{equation*}
\begin{equation*}
H_1^{'(e)}= -\frac{1}{24M_D^2}\int^1_0 dx\,(x-1)^2\left( \frac{(r+x)^2 (x-1)}{\bar{\mathcal{M}}_B^2}+\left(1-x-3 (r+2 x-1) \left(\lambda
   _{\epsilon }+\log \left(\frac{\bar{\mathcal{M}}_B^2}{\bar{\mu}^2}\right)\right)\right) \right),
\end{equation*}
\begin{equation*}
H_2^{(g)}=\frac{1}{18} \int^1_0 dx\,x (x+1) \left(34 x-26+3 (7 x-5) \left(\lambda _{\epsilon }+\log \left(\frac{\bar{\mathcal{M}}_D^2}{\bar{\mu} ^2}\right)\right)\right),
\end{equation*}
\begin{equation*}
H_3^{(g)}=\frac{1}{27} \int^1_0 dx\,x\left(4x(33 - 19 x)-\frac{24 x^2 (x^2 - 1)}{\bar{\mathcal{M}}_D^2}+27 \bar{\mathcal{M}}_D^2+3\left(36-70 x^2+6 x-27\bar{\mathcal{M}}_D^2\right)\left(\lambda
   _{\epsilon }+\log \left(\frac{\bar{\mathcal{M}}_D^2}{\bar{\mu}^2}\right)\right)\right),
\end{equation*}
\begin{equation*}
H_4^{(g)}= \frac{4}{9}\int^1_0 dx\,x\left(\frac{2(x-1) (x+1) x^2}{\bar{\mathcal{M}}_D^2}+\left(9 (x-1) +9 \left(2 x^2-1\right) \left(\lambda _{\epsilon }+\log \left(\frac{\bar{\mathcal{M}}_D^2}{\bar{\mu}
   ^2}\right)\right)\right) \right),
\end{equation*}
\begin{eqnarray*}
&&H_1^{'(g)}=-\frac{1}{216M_D^2}\int^1_0 dx\,x\left(22x(6-5x)+\frac{66x^2(1-x^2)}{\bar{\mathcal{M}}_D^2}+27\bar{\mathcal{M}}_D^2+\right.\\
&&\left.3((6-77 x) x-27\bar{\mathcal{M}}_D^2+36)\left(\lambda _{\epsilon }+\log \left(\frac{\bar{\mathcal{M}}_D^2}{\bar{\mu}^2}\right)\right)\right),
\end{eqnarray*}
\begin{eqnarray*}
&&H_2^{(h)}=\frac{1}{108} \int^1_0 dx\,(1-x)\left(2(x + 1) (x (23 x + 88) - 79)+(139 x-107)\bar{\mathcal{M}}_D^2+\right.\\
&&\left.3\left(5 (x+9) x^2+3 x-37+\left(20 x-16\right)\bar{\mathcal{M}}_D^2\right)\left(\lambda _{\epsilon }+\log \left(\frac{\bar{\mathcal{M}}_D^2}{\bar{\mu}^2}\right)\right)\right),
\end{eqnarray*}
\begin{eqnarray*}
&&H_3^{(h)}=\frac{1}{54} \int^1_0 dx\,(1-x)\left(2(x (x (7 x-195)+281)+3)-\frac{24(x^2-1)^2}{\bar{\mathcal{M}}_D^2}+(47 x-317)\bar{\mathcal{M}}_D^2+\right.\\
&&\left.3\left((x - 93) x^2 - 25 x + 21)+ (4 x + 26)\bar{\mathcal{M}}_D^2 \right)\left(\lambda _{\epsilon }+\log \left(\frac{\bar{\mathcal{M}}_D^2}{\bar{\mu}^2}\right)\right)\right),
\end{eqnarray*}
\begin{eqnarray*}
&&H_4^{(h)}=\frac{2}{81} \int^1_0 dx\,(1-x)^2\left((43 x^2-242x+103)+\frac{3(x^2-1)((x-10) x+1)}{\bar{\mathcal{M}}_D^2}+\right.\\
&&\left.3(x (5 x - 118) - 79)\left(\lambda _{\epsilon }+\log \left(\frac{\bar{\mathcal{M}}_D^2}{\bar{\mu}^2}\right)\right)\right),
\end{eqnarray*}
\begin{eqnarray*}
&&H_1^{'(h)}=-\frac{1}{432M_D^2}\int^1_0 dx\,\left(2(x (x (9 - x (23 x + 21)) + 9) + 26)+\frac{66(x - 1)^3 (x + 1)^2}{\bar{\mathcal{M}}_D^2}+(1-x) (139 x - 395)\bar{\mathcal{M}}_D^2+\right.\\
&&\left.3(1-x)\left(5 (x-5) x^2-109 x+41+(20 x + 14)\bar{\mathcal{M}}_D^2\right)\left(\lambda _{\epsilon }+\log \left(\frac{\bar{\mathcal{M}}_D^2}{\bar{\mu}^2}\right)\right)\right),
\end{eqnarray*}
\begin{equation*}
H_3^{(i,I)}= \int^1_0 dx\,x \bar{\mathcal{M}}_D^2 \left(x+(10-3 x)\left(\lambda _{\epsilon }+\log \left(\frac{\bar{\mathcal{M}}_D^2}{\bar{\mu}^2}\right)\right) -6\right),
\end{equation*}
\begin{equation*}
H_1^{'(i,I)}=-\frac{1}{8M_D^2}H_3^{(i,I)},
\end{equation*}
\begin{equation*}
H_3^{(i,II)}=\frac{2}{9} \int^1_0 dx\,x \bar{\mathcal{M}}_D^2  \left(5 x+6 (5-2 x)\left(\lambda _{\epsilon }+\log \left(\frac{\bar{\mathcal{M}}_D^2}{\bar{\mu}^2}\right)\right)  -38\right),
\end{equation*}
\begin{equation*}
H_1^{'(i,II)}= -\frac{1}{8M_D^2}H_3^{(i,II)}.
\end{equation*}

The functions $f_d^1(\mu^2)$ and $f_d^2(\mu^2)$ used in the regularization of the loop integrals $H_2^{(X)}$ that contribute to the MDMs are
\begin{eqnarray*}
&&f_d^1(\mu)=\frac{1}{9} \left(\left(3 r \left(r \left(r \left(6 r^2+8
   r-7\right)-11\right)+3\right)+34\right) r-3 (r (r (2 r (r (3 r+4)-5)-15)+6)+12) \log \left(\frac{r^2}{\bar{\mu} ^2}\right) r^3+\right.\\
&&\left.3 (r-1) (r+1)^3 (r (2 (r-1) r (3 r+1)+1)+2) \log \left(\frac{1-r^2}{\bar{\mu} ^2}\right)+13\right),\\
&&f_d^2(\mu)=\frac{1}{324}(606\log\left(\bar{\mu}^2\right)-335).
\end{eqnarray*}

\end{document}